\documentclass[prd,aps,showpacs,preprintnumbers,amsmath, amssymb]{revtex4-2}

\usepackage{graphics,epsfig,hyperref}
\usepackage{graphicx}
\usepackage{dcolumn}
\usepackage{bm}
\usepackage{epstopdf}
\usepackage{color}
\usepackage{xcolor}
\usepackage{subfigure}
\usepackage{diagbox} 
\usepackage{adjustbox}
\usepackage{multirow}
\usepackage{amsmath,amssymb,amsfonts}

\usepackage[normalem]{ulem}
\usepackage{xcolor}
\usepackage{mathtools} 
\usepackage{amsthm,amssymb}
\usepackage{mathrsfs}

\begin{document}
\baselineskip=0.8cm
\title{\bf Quasinormal modes and AdS/CFT correspondence of a rotating BTZ-like black hole in the Einstein-bumblebee gravity }
	
\author{Fangli Quan$^{1}$, Zhong-Wu Xia$^{1}$, Rui Ding$^{1}$, Qiyuan Pan$^{1,2}$\footnote{panqiyuan@hunnu.edu.cn} and Jiliang Jing$^{1,2}$\footnote{jljing@hunnu.edu.cn}}
	
\affiliation{
$^1$Department of Physics, Institute of Interdisciplinary Studies, Key Laboratory of Low Dimensional Quantum Structures and Quantum Control of Ministry of Education, Synergetic Innovation Center for Quantum Effects and Applications, and Hunan Research Center of the Basic Discipline for Quantum Effects and Quantum Technologies, Hunan Normal University,  Changsha, Hunan 410081, People's Republic of China} 
\affiliation{$^2$Center for Gravitation and Cosmology, College of Physical Science and Technology, Yangzhou University, Yangzhou 225009, People's Republic of China }
	
\begin{abstract}
\baselineskip=0.6 cm
\begin{center}
{\bf Abstract}
\end{center}

We obtain exact expressions for the quasinormal modes (QNMs) of the massive scalar, fermionic and vector perturbations around a rotating BTZ-like black hole in the Einstein-bumblebee gravity. We find that the Lorentz symmetry breaking (LSB) parameter $\ell$ leaves its imprint only on the imaginary parts of the quasinormal frequencies and the corresponding perturbation field decays more slowly for a larger $\ell$, except for the left-moving quasinormal frequencies $\omega_L$ with positive mass and the right-moving ones $\omega_R$ with negative mass for the fundamental modes under the vector perturbation where the imaginary parts are independent of $\ell$. We also note that, regardless of the kind of perturbations, the real parts depend only on the angular quantum number, which are the same as those in the standard BTZ black hole. Furthermore, we investigate the AdS/CFT correspondence from the QNMs and observe that the expected universal relation for the left and right conformal weights ($h_L,h_R$) of the boundary operators dual to various fields still holds even for the BTZ-like black hole in the Einstein-bumblebee gravity. These results strongly support the AdS/CFT correspondence and could help us better understand the Einstein-bumblebee gravity with the Lorentz symmetry violation.

\end{abstract}
	
\maketitle
\newpage
\vspace*{0.2cm}

\section{Introduction}
Since the advent of gravitational wave astronomy, initiated by the first direct detections by the LIGO/Virgo Collaboration~\cite{LIGOScientific:2016aoc,LIGOScientific:2017vwq,LIGOScientific:2025rsn}, the study of black hole QNMs has gained renewed impetus. In particular, the ringdown phase of a perturbed remnant black hole is well described by a superposition of its QNMs, providing a powerful avenue for precision tests of general relativity (GR) and black hole spectroscopy \cite{Spieksma:2024voy,Berti:2025hly}. QNMs are the characteristic damped oscillations of black holes in response to external perturbations, and encode key information about the underlying spacetime geometry \cite{Regge:1957td, Zerilli:1970wzz, Zerilli:1970se, Kokkotas:1999bd, Nollert:1999ji, Konoplya:2011qq}. In the asymptotically flat spacetime, the analysis of QNMs is typically performed by solving the radial perturbed equations of motion for various spin fields, demanding that the solutions satisfy the physical boundary conditions of being purely ingoing at the horizon and purely outgoing at the infinity \cite{Zhang:2023wwk}. Moreover, this system is intrinsically dissipative, so the frequencies of the QNMs are usually complex. But for a black hole in the asymptotically anti-de Sitter (AdS) spacetime, the ingoing-wave condition still holds at the horizon, while the modified asymptotic condition requires the energy flux to vanish at the AdS boundary \cite{Birmingham:2001pj,Cardoso:2001bb,Chen:2009hg}. 

Over the last few decades, the growing evidence has supported a deep connection between the gravity in AdS spacetime and the conformal field theory (CFT) \cite{Maldacena:1997re,Gubser:1998bc,Witten:1998qj}. According to the AdS/CFT correspondence, the black hole in the bulk of AdS is dual to a thermal state in the CFT on the conformal boundary of the spacetime, and both the decay time of small perturbations and the relaxation time of the state are determined by the imaginary part of dominant QNMs \cite{Maldacena:1997re,Horowitz:1999jd}. A direct computation of the relaxation timescale back to equilibrium is prohibitively difficult \cite{Horowitz:1999jd}, but it becomes accessible through the analysis of the QNMs. We can then understand the properties of the dual CFT, e.g. the conformal dimension associated with the dual operators which is characterised by conformal weights. Following Ref. \cite{Horowitz:1999jd}, for large AdS black holes the imaginary parts of the quasinormal frequencies scale linearly with  the temperature so that the corresponding QNMs describe the decay of perturbations in the dual CFT. The qualitative agreements were found numerically for the Reissner-Nordstr\"{o}m AdS black holes \cite{Wang:2000gsa}, and the remarkable works \cite{Birmingham:2001hc,Birmingham:2001pj} further established a quantitative match for the perturbations  of various spins in the BTZ black hole, in which the quasinormal frequencies correspond exactly to the poles of retarded correlation functions in the dual $(1+1)$-dimensional CFT.

Although GR has made remarkable achievements, ongoing efforts to explore the extensions beyond GR have never ceased \cite{Sotiriou:2008rp,DeFelice:2010aj,Nojiri:2010wj,Capozziello:2011et,Clifton:2011jh,Li:2025gna,Lan:2025cvs,Xie:2025auj,Guo:2024bqe,Cao:2024oud,Li:2025yoz}. As a fundamental symmetry, Lorentz invariance is pivotal in GR, but a number of quantum gravity theories allow its breakdown at high energy \cite{Mattingly:2005re,Jacobson:2005bg,Yang:2023wtu}. This motivates a systematic exploration of the Lorentz symmetry breaking (LSB) in the gravitational sector. Recently, the LSB gravity has gained considerable attention \cite{Media:2024tyg,Laxmi:2021xbi,PriyobartaSingh:2022dis,Media:2022aer,Singh:2024spz,Liu:2024oas,Laxmi:2023yvl,Liu:2024lve}, among which a particularly well-developed theory is the Einstein--bumblebee gravity \cite{Kostelecky:1989jw,Maluf:2020kgf}. The bumblebee model was originally proposed as a string-inspired framework that features the tensor-induced LSB \cite{Kostelecky:1989jw, Kostelecky:1988zi}. As one of the effective field theories, the bumblebee model allows a detailed description of the spontaneous LSB \cite{Escobar:2017fdi}. In this model, the Lorentz violation arises from the dynamics of a vector field known as the bumblebee field $B_\mu$ \cite{Mai:2023ggs,Zhang:2023wwk}. The properties of the bumblebee model in both Riemann and Riemann-Cartan spacetimes have been investigated in Refs. \cite{Bailey:2006fd,Maluf:2013nva,Li:2020dln,Bluhm:2008yt,Escobar:2017fdi,Maluf:2015hda,Hernaski:2014jsa,Carroll:2009em}. Recently Casana \emph{et al.} proposed an exact Schwarzschild-like black hole solution in the Einstein-bumblebee gravity \cite{Casana:2017jkc}, which subsequently gave an impetus to investigating the impact of LSB on this static black hole \cite{Ovgun:2018ran,Kanzi:2019gtu,Oliveira:2021abg,Liu:2022dcn}. Previously, a three-dimensional rotating BTZ-like black hole solution in the Einstein-bumblebee gravity theory has been derived by Ding \emph{et al.} \cite{Ding:2023niy}, then its scalar QNMs spectrum \cite{Chen:2023cjd,Chen:2024pys} and mass ladder operators \cite{Ge:2025xuy} have been investigated.

In this work, we will systematically study the QNMs of various perturbation fields (i.e., scalar, fermionic and vector fields) with the correct boundary conditions around a rotating BTZ-like black hole in the Einstein--bumblebee theory, where the quasinormal frequencies can be derived analytically in all these cases. The motivation for completing this work is two-fold. On one level, it is worthwhile to examine the influence of the LSB on the QNMs of fileds with various spins and further understand the LSB in the Einstein-bumblebee gravity. On another more speculative level, it would be important to investigate the AdS/CFT correspondence from the QNMs, which will give the left and right conformal weights ($h_L,h_R$) of the boundary operators dual to perturbation fields of different spins. According to the AdS/CFT correspondence, the spin-$s$ field propagating in standard AdS${}_3$ is related to the operator in the dual CFT \cite{Aharony:1999ti}
\begin{eqnarray}\label{HspinLR}
	h_R + h_L = \Delta,\quad h_R - h_L = \pm s.
\end{eqnarray}
Thus, it is of interest to examine whether the relation (\ref{HspinLR})  still holds for the rotating BTZ-like black hole in the Einstein-bumblebee gravity and whether the LSB parameter affects the conformal dimension $\Delta$ of operators in the CFT. Moreover, we also explore the AdS/CFT correspondence under different perturbations, which provides an accessible approach to computing the relaxation time of the system toward the thermal equilibrium.

The structure of this work is organized as follows. In Section II, we briefly review the rotating BTZ-like black hole in the Einstein-bumblebee gravity. Then, in Section III we analytically calculate the QNMs under the scalar perturbation, study the AdS/CFT correspondence from the QNMs, and analyze the left and right conformal weights $h_L$ and $h_R$ of the operators dual to the scalar field at the AdS boundary. We extend the investigation to the fermionic perturbation in Section IV and the vector perturbation in Section V. Finally, in Section VI we will include our conclusions.

\section{Background geometry}
We begin with the action of the Einstein-bumblebee gravity in three dimensions \cite{Bluhm:2004ep,Bertolami:2005bh,Kostelecky:2008in,Seifert:2009gi,Maluf:2014dpa,Paramos:2014mda,Assuncao:2019azw,Capelo:2015ipa} 
\begin{eqnarray}\label{action}
S=\int d^3x \sqrt{-g}\left[\frac{R-2\Lambda}{2\kappa} + \frac{\xi}{2\kappa} B^{\mu} B^{\nu}R_{\mu \nu}-\frac{1}{4} B^{\mu\nu} B_{\mu\nu} - V\left( B_{\mu} B^{\mu} \pm b^2 \right)\right],
\end{eqnarray}
where $R$ describes the Ricci scalar, $\Lambda=-1/l^2$ represents a negative cosmological constant related to the AdS radius $l$, $\kappa =8\pi G/c^3$ denotes a constant associated with the three-dimensional Newtonian constant $G$, $\xi$ is the coupling constant between the gravity and the bumblebee vector field $B_{\mu}$ with the strength $B_{\mu\nu}=\partial_{\mu}B_{\nu}-\partial_{\nu}B_{\mu}$, $b$ stands for a real positive constant, and the signs ``$\pm$" mean that $b_{\mu}$ is timelike or spacelike. The potential $V$ is chosen so that its minimum enforces a nonzero vacuum expectation value $\langle B_{\mu}\rangle=b_{\mu}$ with $b_{\mu}b^{\mu}=\mp b^2$ for the bumblebee field, and the nontrivial vacuum expectation value selects a prior direction, leading to the violation of local Lorentz invariance. Varying the action (\ref{action}) yields a modified Einstein equation, and the corresponding rotating BTZ-like black hole solution is expressed as \cite{Ding:2023niy}
\begin{equation}\label{spacetime}
ds^2 = -f(r) dt^2 +\frac{1+\ell}{f(r)} dr^2 + r^2 \left(d\phi - \frac{j} {2r^2}dt\right)^2,
\end{equation}
where
\begin{equation}\label{metricfr}
f(r)=\frac{r^2}{l^2} - M +\frac{j^2}{4r^2},
\end{equation}
with the mass parameter $M$ and rotation parameter $j$ of the black hole, and the LSB parameter $\ell=\xi b^{2}$. One can observe that the spacetime (\ref{spacetime}) is singular when $\ell=-1$ according to the determinant of metric $g=-(1+\ell)r^2$ and the Kretschmann scalar $R_{\mu \nu \rho \lambda}=12/[l^4 (1+\ell)^2]$. In order to keep the spacetime free of the curvature singularity, we impose the constraint $\ell>-1$. The black hole solution (\ref{spacetime}) has an outer horizon and an inner horizon
\begin{equation}
r^{2}_{\pm} = \frac{l^2}{2}\left(M \pm \sqrt{M^2 -\frac{j^2}{l^2}}\right),
\end{equation} 
which are independent of the LSB parameter $\ell$. The temperature of the outer horizon for this rotating BTZ-like black hole is \cite{Ding:2023niy}
\begin{equation}\label{hawkingt}
T_{+}=\frac{2\;T_{L} T_{R}}{T_{L}+T_{R}}, 
\end{equation}
with the temperatures of left- and right-moving sectors at thermal equilibrium of the black hole
\begin{eqnarray}\label{HawkingTLR}
T_L=\frac{r_+ - r_-}{2\pi l^2\sqrt{1+\ell}},\quad T_R=\frac{r_+ + r_-}{2\pi l^2\sqrt{1+\ell}}.
\end{eqnarray}

For convenience, in the following we will introduce the coordinates $~ x^{\mu} = (x^+, \rho, x^-)$ which are related to the Schwarzschild-like coordinates $(t, r, \phi)$ through the transformation~\cite{Birmingham:2001pj}
\begin{equation} \label{transformations}
x^+ =r_+ t - r_- \phi,  \quad  x^- = r_+ \phi - r_-   t,  \quad \tanh \rho = \sqrt{\frac{r^2 - r_+^2}{r^2 - r_-^2}}.
\end{equation}
Thus, the form of the metric \eqref{spacetime} becomes
\begin{equation}\label{metric}
ds^2=-\sinh^2 \rho ~ {(d x^+)}^2 + (1+\ell) d \rho^2 + \cosh^2 \rho ~ {( d x^-)}^2,
\end{equation}
where the AdS radius is set to unity, i.e., $l=1$.

\section{Scalar QNMs and the AdS/CFT correspondence}
The scalar perturbations are described by the Klein-Gordon equation~\cite{Konoplya:2011qq}
\begin{equation}\label{Klein-GordonEQ}
\frac{1}{\sqrt{-g}}\frac{\partial}{\partial x^{\mu}}\left( \sqrt{-g} g^{\mu\nu}
\frac{\partial}{\partial x^{\nu}}\right) \Phi -m ^2 \Phi= 0,
\end{equation}
with the mass of the scalar field $m$. To separate the variables, we use the following ansatz~\cite{Chen:2023cjd}
\begin{equation}\label{s2}
\Phi(x^+, \rho, x^-)=e^{-i(k_+x^++k_-x^-)}R(\rho),
\end{equation}
where
\begin{eqnarray}\label{s3}
(k_+ +k_-)(r_+-r_-)=\omega-k,\qquad 
(k_+ -k_-)(r_++r_-)=\omega+k.	
\end{eqnarray}
Here, $\omega$ and $k$ are the energy and angular momentum of the scalar fields. Thus, the radial field equation is
\begin{equation}\label{kgeq}
\frac{\mathrm{d} ^2 R(\rho)}{\mathrm{d} \rho^2} +\left(\coth\rho+\tanh\rho\right)\frac{\mathrm{d} R(\rho)}{\mathrm{d} \rho} +(1+\ell)\left(-k^2 \mathrm{sech}^2\rho +\omega^2\mathrm{csch}^2\rho -m^2 \right) R(\rho) =0.
\end{equation}
For simplicity, we introduce a new coordinate transformation $z=\tanh^2\rho $~\cite{Birmingham:2001pj}. Under this transformation, Eq. \eqref{kgeq} takes the form 
\begin{eqnarray}\label{eq:radial}
z(1-z)\frac{\mathrm{d}^2R(z)}{\mathrm{d} z^2}+(1-z)\frac{\mathrm{d}
R(z)}{\mathrm{d} z}+\left[\frac{k_+^2(1+\ell)}{4z}
-\frac{k_-^2(1+\ell)}{4}-\frac{m^2(1+\ell)}{4(1-z)}\right]R(z)=0.	
\end{eqnarray}
Redefining the radial function as $R(z)=z^{\alpha_s}(1-z)^{\beta_s} F(z)$, Eq. \eqref{eq:radial} can be rewritten as the hypergeometric differential equation~\cite{olver2010nist}
\begin{equation}
z(1-z)\frac{d^2F(z)}{dz^2} +\left[c_s -(1+a_s+b_s)z\right]\frac{dF(z)}{dz} -a_s b_s F(z)=0,
\end{equation}
with
\begin{eqnarray}\label{abc}
\alpha_s =&&-\frac{ik_+\sqrt{1+\ell}}{2},
\qquad \beta_s = \frac{1}{2}\left[1-\sqrt{1+m^2(1+\ell)}\right],\notag\\
a_s=&&\frac{(k_+-k_-)\sqrt{1+\ell}}{2i}+\beta_s,\qquad b_s= \frac{(k_++k_-)\sqrt{1+\ell}}{2i}+\beta_s,
\qquad c_s= 1+2\alpha_s, 
\end{eqnarray}	
which implies a modified Breitenlohner-Freedman bound~\cite{Breitenlohner:1982jf}, $m^2\ge- 1/(1+\ell)$. 

\begin{figure}[htbp]
\subfigure{\includegraphics[scale=0.55]{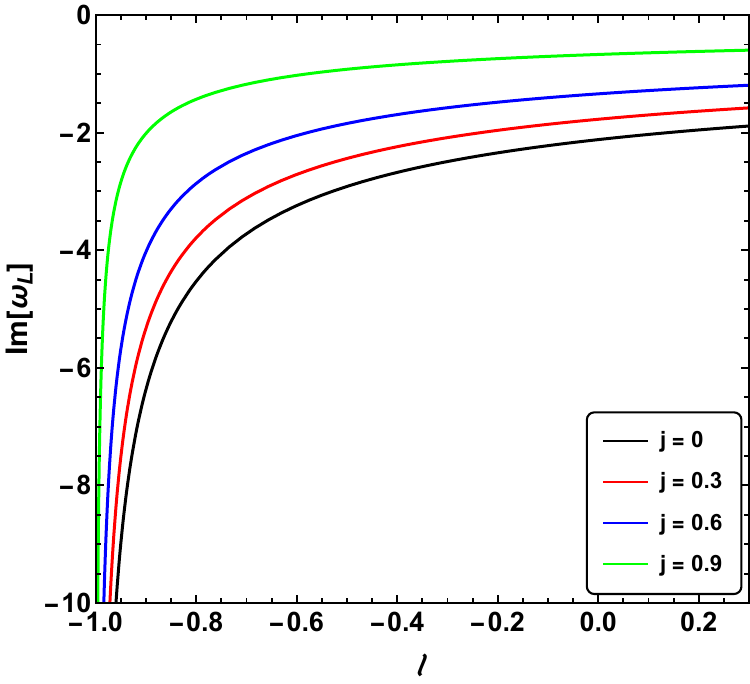}}\hspace{8ex}
\subfigure{\includegraphics[scale=0.55]{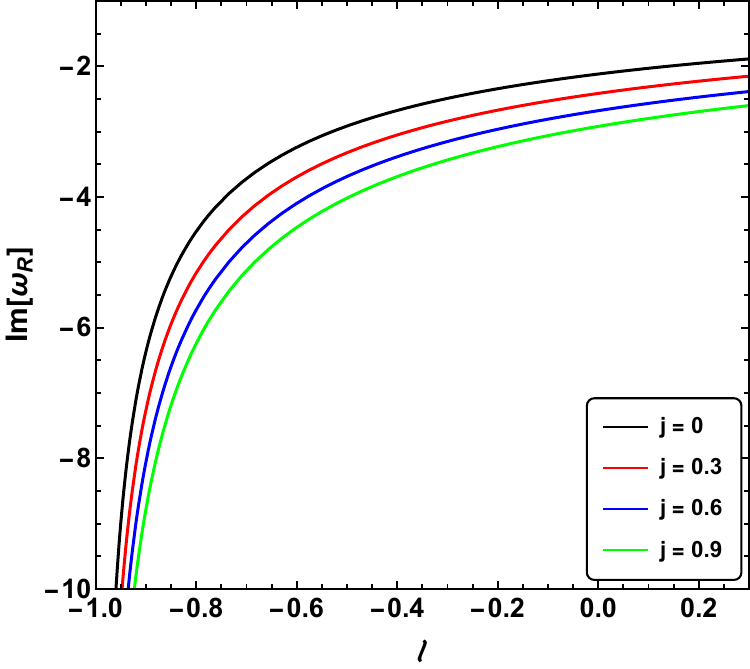}}\\ \vspace*{4ex}
\subfigure{\includegraphics[scale=0.55]{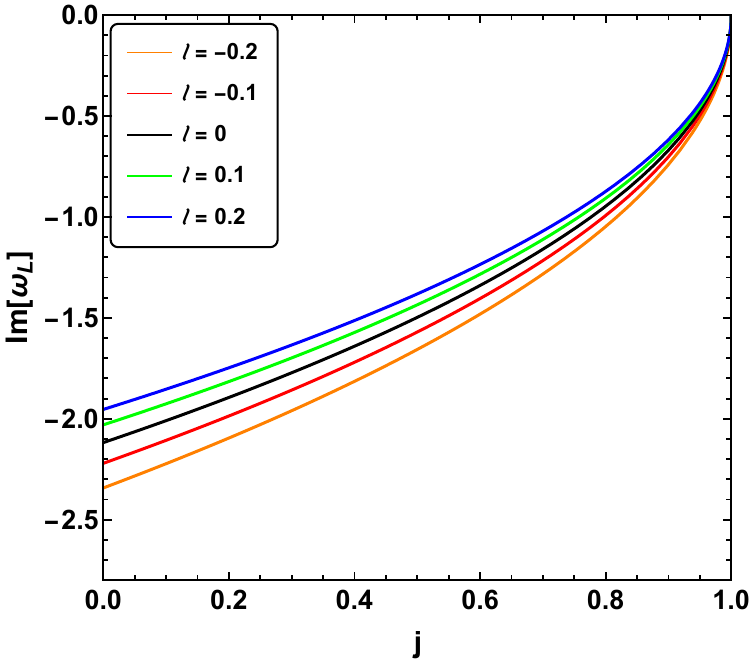}} \hspace{8ex}
\subfigure{\includegraphics[scale=0.55]{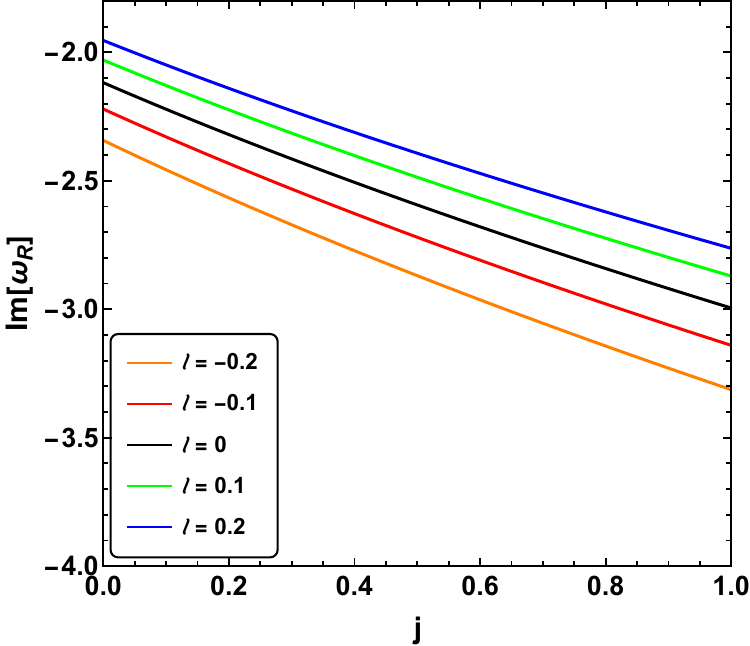}}
\caption{ Variation of the imaginary parts of left- and right-moving quasinormal frequencies for the fundamental modes ($n=0$) under the scalar perturbation with the LSB parameter $\ell$ and the rotation parameter $j$. Here we set $M=1$ and $m=0.5$.}\label{scalarQNMs_s}
\end{figure}

In order to obtain the scalar QNMs in the BTZ-like black hole, at the outer horizon ($z=0$) we impose an ingoing-wave boundary condition 
\begin{equation}
R(z)= A\; z^{\alpha_s} (1-z)^{\beta_s} F(a_s,b_s;c_s;z),
\end{equation}
where $A$ is an integration constant. To connect the wave functions at the outer horizon ($z=0$) and at the spatial infinity ($z \to 1$), we make use of Kummer's formula
\begin{eqnarray}\label{transform}
F(a,b;c;z)&&=\frac{\Gamma(c)\Gamma(c-a-b)}{\Gamma(c-a)\Gamma(c-b)}F(a,b;a+b-c+1;1-z) \notag\\
&&+(1-z)^{c-a-b}\frac{\Gamma(c)\Gamma(a+b-c)}{\Gamma(a)\Gamma(b)}F(c-a,c-b;c-a-b+1;1-z).
\end{eqnarray}
At the spatial infinity ($z \to 1$), we have to impose the appropriate boundary conditions on the radial equation. It should be noted that the QNMs for the scalar field were already found in \cite{Chen:2023cjd} by imposing the Dirichlet boundary condition. Here, we compute these modes by employing the vanishing energy flux
\begin{eqnarray}\label{s8}
{\cal F}=\sqrt{-g}\frac{1}{2i}\left(R^*\partial_\rho R
- R\partial_{\rho}R^{*}\right).
\end{eqnarray}
For $m^2>-1/(1+\ell)$, the leading divergent term of the energy flux is \cite{olver2010nist}
\begin{equation}\label{s10}
\left|\frac{\Gamma(c_s)\Gamma(c_s-a_s -b_s)}{\Gamma(c_s-a_s )\Gamma(c_s-b_s)}\right|^2(1-z)^{2\beta_s}.
\end{equation}
Thus, the vanishing asymptotic flux boundary condition leads to $c_s-a_s=-n$ or $c_s-b_s=-n $, i.e., 
\begin{equation}\label{qnms}
\frac{i}{2}(k_+\pm k_-)\sqrt{1+\ell}=n+\frac{1}{2} \left[1+\sqrt{1+m^2(1+\ell)}\right],
\end{equation}
where $n$ is a non-negative integer. By solving Eq. \eqref{qnms}, we can obtain both the left- and right-moving quasinormal frequencies
\begin{eqnarray}\label{qnmsDRL}
\omega_L&&=k-\frac{2i(r_+-r_-)}{\sqrt{1+\ell}}\left\{n+\frac{1}{2}\left[1+\sqrt{1+m^2(1+\ell)}\right]\right\}, \\
\omega_R&&=-k-\frac{2i(r_++r_-)}{\sqrt{1+\ell}}\left\{n+\frac{1}{2}\left[1+\sqrt{1+m^2(1+\ell)}\right]\right\}.
\end{eqnarray}
However, it is also possible to find a second set of quasinormal frequencies for $-1/(1+\ell)<m^2<0$, where we have the relation $ a_s=-n$ or $ b_s=-n $, i.e.,
\begin{eqnarray}
\omega_L&&=k-\frac{2i(r_+-r_-)}{\sqrt{1+\ell}}\left\{n+\frac{1}{2}\left[1-\sqrt{1+m^2(1+\ell)}\right]\right\},\\
\omega_R&&=-k-\frac{2i(r_++r_-)}{\sqrt{1+\ell}}\left\{n+\frac{1}{2}\left[1-\sqrt{1+m^2(1+\ell)}\right]\right\}.\label{qnmsNRL}
\end{eqnarray}
Obviously, we can recover the scalar quasinormal frequencies of the BTZ black hole by taking $\ell=0$ \cite{Ferreira:2017cta,Dappiaggi:2017pbe}. It is interesting to note that, from Eq. (\ref{qnmsDRL}) to Eq. (\ref{qnmsNRL}), only the imaginary parts of the quasinormal frequencies depend on the LSB parameter $\ell$ and the rotation parameter $j$. In Fig. \ref{scalarQNMs_s}, we present the impacts of $\ell$ and $j$ on imaginary parts of the fundamental quasinormal frequencies ($n=0$). Here, for concreteness we take $m=0.5$ in the following analysis. It is shown clearly that, regardless of the left-moving quasinormal frequencies or right-moving ones, the imaginary parts increase with increasing $\ell$, which means that the scalar field decays more rapidly for smaller $\ell$. Moreover, as $j$ increases, we find that the imaginary parts increase for the left-moving modes but decrease for the right-moving modes, which indicates that increasing the rotation parameter $j$ leads to different behaviors in the damping time scale for the left and right branches whereas the oscillation time scale remains unaffected. Similar results can be obtained for higher overtone modes.
	
On the other hand, we observe that the imaginary parts of the quasinormal frequencies scale linearly with the left- and right-moving temperatures by comparing Eq. (\ref{HawkingTLR}) with Eqs. (\ref{qnmsDRL})---(\ref{qnmsNRL}). Note that, corresponding to the poles of the retarded correlation function of the perturbation operator in CFT, the QNMs under various perturbations can be rewritten as~\cite{Birmingham:2001pj}
\begin{eqnarray}\label{QNMsCFT}
\omega_L = k - 4\pi i T_L(n + h_L),\qquad\omega_R = -k - 4\pi i T_R(n + h_R),
\end{eqnarray}
where $n$ is a non-negative integer and $h_{L,R}$ is the conformal weight of the operator in the dual CFT which contains both left- and right-moving sectors \cite{Brown:1986nw}. Thus, for the exact QNM spectrum of the scalar perturbation, we obtain the left and right conformal weights for $m^2>-1/(1+\ell)$
\begin{eqnarray}\label{HscalarDLR}
h_R^s = h_L^s =\frac{1}{2}\left[1+\sqrt{1+m^2(1+\ell)}\right],
\end{eqnarray}
and for $-1/(1+\ell)<m^2<0$
\begin{eqnarray}\label{HscalarNLR}
h_R^s = h_L^s =\frac{1}{2}\left[1-\sqrt{1+m^2(1+\ell)}\right].
\end{eqnarray}
For the scalar field ($s=0$) considered here, from Eqs. (\ref{HspinLR}), (\ref{HscalarDLR}) and (\ref{HscalarNLR}) we have $\Delta = 1 \pm \sqrt{1 + m^2(1+\ell)}$ and $h_R^s-h_L^s=0$, regardless of $m^2>-1/(1+\ell)$ or $-1/(1+\ell)<m^2<0$. Obviously, the relation (\ref{HspinLR}) still holds even for the BTZ-like black hole in the Einstein-bumblebee gravity. Moreover, for the perturbed CFT thermal system which returns to equilibrium with a characteristic relaxation time given by the imaginary part of the lowest quasinormal frequency, i.e., $\tau=1/|\mathrm{Im}[\omega]|$, we find that the relaxation time  $\tau$ increases as the LSB parameter $\ell$ increases both for the left- and right-moving modes, but exhibits opposite behaviors for the left- and right-moving modes as the rotation parameter $j$ increases.

\section{Fermionic QNMs and the AdS/CFT correspondence}
The QNMs of a fermionic perturbation for a three-dimensional background spacetime can be obtained by solving the Dirac equation~\cite{Gonzalez:2014voa}
\begin{eqnarray}\label{diraceq}
(\gamma^{\mu} \bigtriangledown_{\mu} +m) \Psi=0,
\end{eqnarray}
with the covariant derivative
\begin{eqnarray}
\bigtriangledown_{\mu}=\partial_{\mu}+\frac{1}{2}\omega_{\mu}{}^{ab} J_{a b},
\end{eqnarray}	
where the spin connection of tetrad $\omega_{\mu}{}^{ab}$ is defined as
\begin{equation}
\omega_{\mu}{}^{ab}
= e^{a}{}_{\nu}\,\eta^{bc}\,\partial_{\mu} e_{c}{}^{\nu}
+e^{a}{}_{\nu}\,\eta^{bc}\,e_{c}{}^{\lambda}\,\Gamma^{\nu}{}_{\mu\lambda}, 
\end{equation}
with the affine connection $\Gamma^{\nu}{}_{\mu\lambda}$. And $J_{a b}$ are the generators of Lorentz group
\begin{eqnarray}
J_{a b}=\frac{1}{4}\left[\gamma_a,\gamma_b\right],
\end{eqnarray}
where $\gamma^a$ are the gamma matrices in the flat spacetime
\begin{equation} \label{gammarep}
	\gamma^0 = i \sigma_2 =
	\left( \begin{array}{ccccc}
		0  & 1  \\
		-1   & 0  \\ 
	\end{array} \right),  \quad \quad 
	\gamma^1 =  \sigma_1 =
	\left( \begin{array}{ccccc}
		0  & 1  \\
		1   & 0  \\  
	\end{array} \right), \quad \quad 
	\gamma^2 =  \sigma_3 =
	\left( \begin{array}{ccccc}
		1  & 0  \\
		0   & -1  \\ 
	\end{array} \right),  
\end{equation} 
which satisfy $\{ \gamma^a, \gamma^b \} = 2 \eta_{ab}$. Here $\sigma_i ~ (i=1,2,3)$ are the Pauli matrices.

For the rotating BTZ-like black hole (\ref{metric}) in the Einstein-bumblebee gravity, by introducing the tetrads $e^a_{\ \mu} = \text{diag}(\sinh \rho,\ \sqrt{1+\ell},\ \cosh \rho) $, the Dirac gamma matrices in this curved spacetime $\gamma^{\mu}$ are related with  $\gamma^a$ via the tetrads as $\gamma^{\mu}=\gamma^{a} e_a^{~~\mu}$. Based on the above definitions, the Dirac equation in this BTZ-like black hole takes the form
\begin{equation} \label{diracnovo2}
\bigg[\frac{\gamma^1}{\sqrt{1+\ell}} \bigg( \frac{\partial}{\partial \rho}  + \frac{\cosh \rho}{2\sinh \rho} + \frac{\sinh \rho}{2\cosh \rho} \bigg)  + \gamma^0 \frac{1}{\sinh \rho}   
\frac{\partial}{\partial x^+} 
+  \gamma^2 \frac{1}{\cosh \rho} \frac{\partial}{\partial x^-} + m \bigg] \Psi=0.
\end{equation}
In order to solve this equation for the wavefunction $ \Psi=\Psi(x^+,\rho, x^-) $, we use the ansatz~\cite{Birmingham:2001pj,Dasgupta:1998jg}
\begin{equation} \label{anzatz}
\Psi =\left(\begin{matrix} \psi_1  \\ \psi_2  \\   \end{matrix}\right)
=\left(\begin{matrix} \psi_1 (r) \\ \psi_2 (r) \\ \end{matrix}\right)\text{exp}\left[-i(\omega t - k \phi)\right]
= \left(\begin{matrix} \psi_1 (\rho) \\ \psi_2 (\rho) \\ \end{matrix}\right)\text{exp}\left[-i(k_+ x^+ +k_- x^-)\right],
\end{equation}
where $\omega$ and $k$ are related to $k_+$ and $k_-$ by Eq.~\eqref{s3}, and $(1,2)$ refer to the two components of the spinor. For simplicity, we define a new set of wavefunctions $\psi_{1,2}'$ as
\begin{eqnarray}\label{f1}
\psi_1(\rho)\pm\psi_2(\rho)&=&\sqrt{\frac{\cosh \rho \pm \sinh\rho}{\cosh \rho \sinh \rho}} \left[\psi_1'(z)\pm\psi_2'(z)\right],
\end{eqnarray}
 with $z =\tanh^2\rho$. So the Dirac equation \eqref{diracnovo2} transforms into 
\begin{eqnarray}\label{f4}
&&2(1-z)z^{1/2}\frac{\mathrm{d} \psi_1'}{\mathrm{d} z}+ i\sqrt{1+\ell}\left( \,k_+ z^{-1/2} + \, k_- z^{1/2}\right)\psi_1' + \left[i\,(k_+ + k_-)\sqrt{1+\ell} +m  \sqrt{1+\ell} + \frac{1}{2}\right]\psi_2' = 0,  \notag\\ 
&&2(1-z)z^{1/2}\frac{\mathrm{d} \psi_2'}{\mathrm{d}z} - i\sqrt{1+\ell}\left(\, k_+ z^{-1/2} + \, k_- z^{1/2}\right)\psi_2' - \left[i(k_+ + k_-)\sqrt{1+\ell}- m\sqrt{1+\ell} - \frac{1}{2}\right]\psi_1' = 0.
\end{eqnarray}
Separating the equations for $\psi_1'$ and $\psi_2'$ and choosing the following transformation
\begin{equation}
\psi_i'~=~ B_i\; z^{\alpha_i} (1-z)^{\beta_i} F_i (z),
\end{equation}
where $B_i$ are integration constants with $i=1,2$, we finally obtain the following hypergeometric equations
\begin{eqnarray}
&&z(1-z) \frac{\mathrm{d} ^2F_i}{\mathrm{d}z^2} + \left[ \left(2\alpha_i+\frac{1}{2}\right) - \left(2\alpha_i + 2\beta_i +\frac{3}{2}\right)z\right]\frac{\mathrm{d} F_i}{dz} \notag \\
&& -\left[\alpha_i\left(\alpha_i+\frac{1}{2}\right) +\beta_i\left(\beta_i+\frac{1}{2}\right) + 2\alpha_i\beta_i - \frac{i k_-\sqrt{1+\ell} - k_-^2(1+\ell)}{4}\right] F_i =0.
\end{eqnarray}
	
\begin{figure}[htbp]
	\subfigure{\includegraphics[scale=0.55]{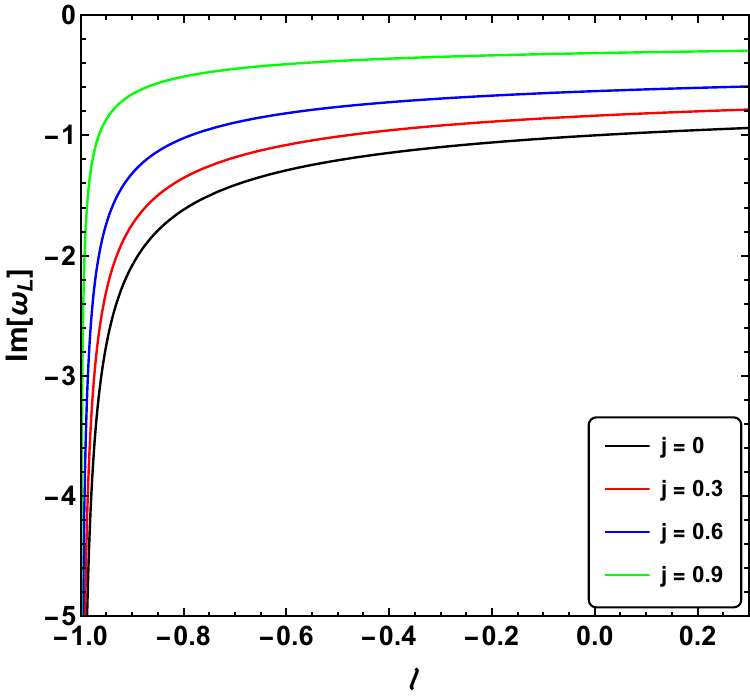}}\hspace{8ex}
	\subfigure{\includegraphics[scale=0.55]{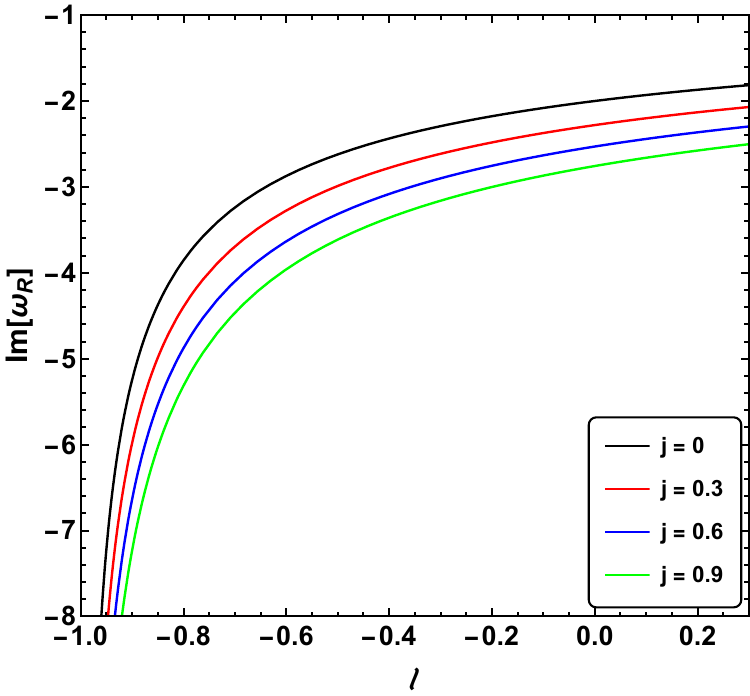}}\\ \vspace*{4ex}
	\subfigure{\includegraphics[scale=0.55]{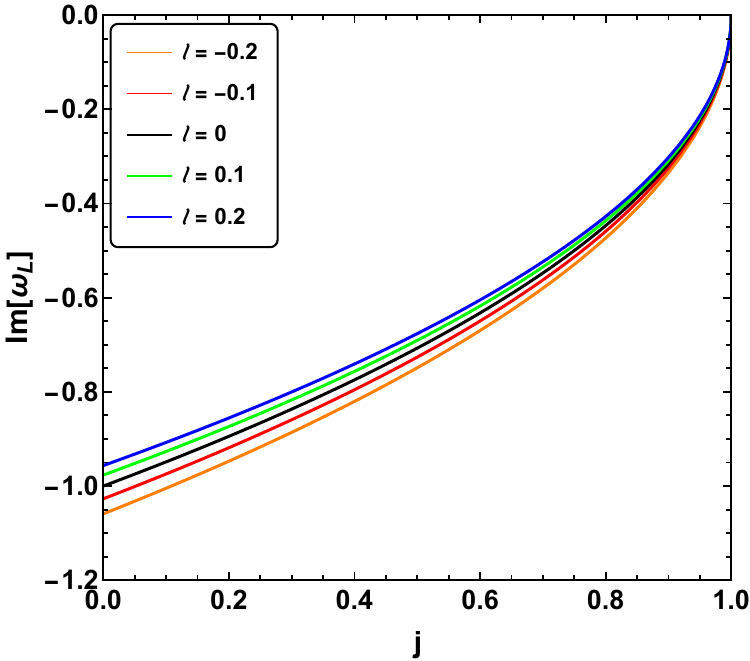}} \hspace{8ex}
	\subfigure{\includegraphics[scale=0.55]{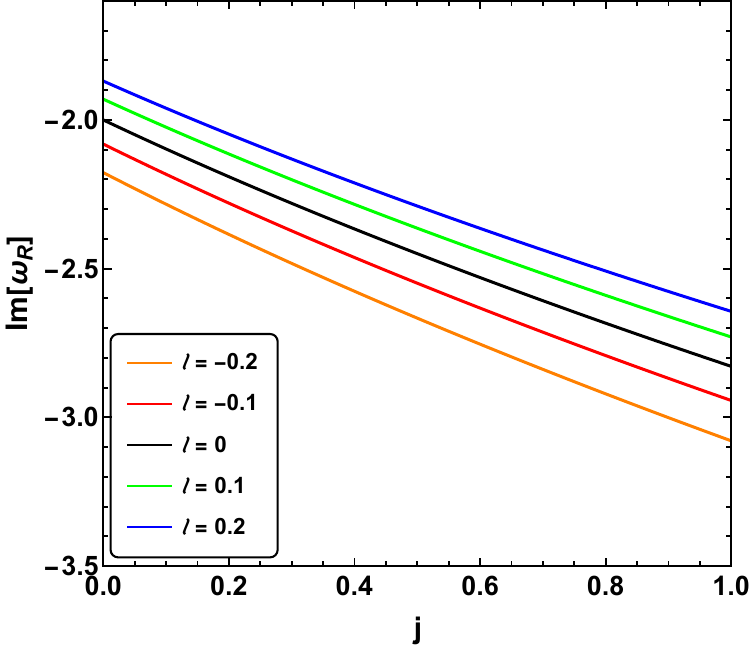}}
	\caption{ Variation of the imaginary parts of left- and right-moving quasinormal frequencies for the fundamental modes ($n=0$) under the fermionic perturbation with the LSB parameter $\ell$ and the rotation parameter $j$. Here we set $M=1$ and $m=0.5$.}\label{diracQNMs_s}
	\end{figure}

At the horizon ($z = 0$), the ingoing-wave solutions are
\begin{eqnarray}\label{p1}
\psi_1'&=&z^{\alpha_f} (1-z)^{\beta_f} F(a_f,b_f;c_f;z) ,\notag\\
\psi_2'&=&\left({\frac{a_f-c_f}{c_f}}\right) z^{\alpha_f + 1/2}(1-z)^{\beta_f} F(a_f,b_f+1;c_f+1;z),
\end{eqnarray}
with
\begin{eqnarray}
&&  \alpha_f=-\frac{i k_+\sqrt{1+\ell}}{2}, \quad \beta_f =-\frac{1}{2}\left(\frac{1}{2}+m\sqrt{1+\ell} \right), \quad c_f = \frac{1}{2}+2\alpha_f, \notag \\
&& a_f = \frac{1}{2} + \alpha_f + \beta_f + \frac{ik_-\sqrt{1+\ell}}{2},\quad b_f =  \alpha_f + \beta_f - \frac{ik_-\sqrt{1+\ell}}{2}, \quad \frac{B_2}{B_1}= \frac{a_f-c_f}{c_f}. 
\end{eqnarray}  
The asymptotic form of $\psi_1$ and $\psi_2$ can be determined by the connection formula Eq. \eqref{transform}. In analogy with the scalar perturbation, we should impose the following condition  of the energy flux~\cite{Das:1999pt}
\begin{equation}
{\cal F}=\sqrt{-g}\bar\Psi e^{\rho}_1 \gamma^1 \Psi\simeq(1-z)^{-1}(|\psi_1|^2-|\psi_2|^2)=0
\end{equation}
at the infinity, which results in the emergence of fermionic QNMs. For $m>- 1/(2\sqrt{1+\ell})$, after expanding the hypergeometric function, we find that the vanishing energy flux is equivalent to the following condition
\begin{equation}
{ \frac{\Gamma (c_f)\Gamma (c_f-a_f-b_f)}{\Gamma
(c_f-a_f)\Gamma (c_f-b_f)}} =0,
\end{equation}
leading to
\begin{eqnarray}\label{diracqnms}
\frac{i}{2}(k_++k_-)\sqrt{1+\ell} =n+\frac{1}{4}+\frac{m\sqrt{1+\ell}}{2},
\end{eqnarray}
or
\begin{eqnarray}
\frac{i}{2}(k_+-k_-)\sqrt{1+\ell} =n+\frac{3}{4}+\frac{m\sqrt{1+\ell}}{2}.
\end{eqnarray}
Thus, the left- and right-moving quasinormal frequencies $\omega_L$ and $\omega_R$ for the fermionic perturbation are
\begin{eqnarray}\label{diracDLR}
\omega_L&&=k-\frac{2i(r_+-r_-)}{\sqrt{1+\ell}}\left(n+\frac{1}{4}+\frac{m\sqrt{1+\ell}}{2}\right),\\
\omega_R&&=-k-\frac{2i(r_++r_-)}{\sqrt{1+\ell}}\left(n+\frac{3}{4}+\frac{m\sqrt{1+\ell}}{2}\right).
\end{eqnarray}
Similarly, for $m<- 1/(2\sqrt{1+\ell})$, we have
\begin{eqnarray}
\omega_L&&=k-\frac{2i(r_+-r_-)}{\sqrt{1+\ell}}\left(n+\frac{1}{4}-\frac{m\sqrt{1+\ell} }{2}\right),\label{diracL} \\ 
\omega_R&&=-k-\frac{2i(r_++r_-)}{\sqrt{1+\ell}}\left(n+\frac{3}{4}-\frac{m\sqrt{1+\ell} }{2}\right).\label{diracNLR}
\end{eqnarray}
In the limit $\ell \to 0$, these results reduce to the fermionic QNMs in the BTZ black hole \cite{Birmingham:2001pj}. Obviously, similar to the scalar QNMs obtained in previous section, the real parts $\text{Re}[\omega_{L,R}]$ of fermionic quasinormal frequencies are only determined by the angular quantum number $k$ but the imaginary parts $\text{Im}[\omega_{L,R}]$ depend on the LSB parameter $\ell$ and the rotation parameter $j$. In Fig.~\ref{diracQNMs_s}, we present how the imaginary parts of the fundamental quasinormal frequencies vary with the parameters $\ell$ and $j$ for the fermionic perturbation, which shows that varying $j$ and $\ell$ significantly affect $\mathrm{Im}[\omega_{L,R}]$. Specifically, for both the left and right branches, increasing $\ell$ consistently increases $\mathrm{Im}[\omega_{L,R}]$, indicating that a positive $\ell$ makes the fermionic perturbations of the spacetime decay faster. Furthermore, as $j$ increases, $\mathrm{Im}[\omega_L]$ increases whereas $\mathrm{Im}[\omega_R]$ decreases, implying that a larger $j$ leads to a slower decay of the fermionic field in the left-moving branch but a faster decay in the right-moving branch. So we conclude that the dependence of imaginary parts for the fermionic QNMs on the LSB parameter $\ell$ and the rotation parameter $j$ is similar to that of imaginary parts for the scalar QNMs. Similar results can be extended to higher overtone modes.

To connect the above results with CFT prediction, we will give the conformal weights of this massive fermionic perturbation. According to Eq. (\ref{QNMsCFT}) and Eqs. (\ref{diracDLR})---(\ref{diracNLR}), for $m>- 1/(2\sqrt{1+\ell})$, we obtain
\begin{eqnarray}
&&h_R^f = \frac{3}{4} + \frac{1}{2}m\sqrt{1+\ell},\qquad
h_L^f = \frac{1}{4} + \frac{1}{2}m\sqrt{1+\ell}, \notag\\
&&h_R^f + h_L^f = 1+m\sqrt{1+\ell} ,\qquad
h_R^f - h_L^f = \frac{1}{2},
\end{eqnarray}
and for $m<- 1/(2\sqrt{1+\ell})$
\begin{eqnarray}
&&h_R^f = \frac{1}{4} - \frac{1}{2}m\sqrt{1+\ell},\qquad
h_L^f = \frac{3}{4} - \frac{1}{2}m\sqrt{1+\ell}, \notag\\
&&h_R^f + h_L^f = 1-m\sqrt{1+\ell} ,\qquad
h_R^f - h_L^f = -\frac{1}{2}.
\end{eqnarray}
From Eq. (\ref{HspinLR}), we have the conformal dimension $\Delta=1\pm m\sqrt{1+\ell}$ for fermionic fields, which depends only on the LSB parameter $\ell$. It should be noted that Eq. (\ref{HspinLR}) still holds for a fermionic perturbation around the rotating BTZ-like black hole in the Einstein-bumblebee gravity. In addition, using the AdS/CFT dictionary, we observe that the characteristic relaxation time  $\tau$ increases as the LSB parameter $\ell$ increases both for the left- and right-moving modes for the fermionic perturbation, which indicates that it will take more time for the perturbation to return to thermal equilibrium for a larger $\ell$.

\section{Vector QNMs and the AdS/CFT correspondence}	
In this section we consider the vector perturbation which is governed by the massive Maxwell equation in the first order \cite{Birmingham:2001pj}
\begin{equation}\label{v2}
\epsilon_{\lambda}^{\alpha\beta}\partial_{\alpha}A_{\beta}=-mA_{\lambda},
\end{equation}
where $\epsilon_{\lambda}^{\ \alpha\beta}$ is the Levi-Civita tensor with $\epsilon^{x^+\,\rho\,x^- }=1/\sqrt{-g}$. As a matter of fact, the solutions to this equation coincide with the second-order ordinary differential equation for the massive vector field in three-dimensional spacetime \cite{Chen:2009hg}
\begin{equation}
\triangledown _{\mu}F^{\mu\nu}=m^2A^{\nu},
\end{equation}
where $F_{\mu\nu}=A_{\nu,\mu}-A_{\mu,\nu}$ with the vector potential $A_{\mu}$. For the rotating BTZ-like black hole (\ref{metric}) in the Einstein-bumblebee gravity, it is convenient to introduce the following ansatz
\begin{equation}\label{v3}
A_\mu=e^{-i(k_+x^++k_-x^-)}A_\mu(\rho),
\end{equation}
which results in the equations of motion
\begin{eqnarray}
&&\tanh\rho\left(\frac{\mathrm{d} A_{\rho}}{dx^-}-\frac{ \mathrm{d} A_{-}}{d\rho}\right) +m\sqrt{1+\ell}  A_+=0,\label{A1} \\ 
&&(1+\ell)\,\mathrm{csch}\rho\,\mathrm{sech}\rho\left(\frac{\mathrm{d} A_{+}}{dx^-}-\frac{\mathrm{d} A_{-}}{dx^+}\right) +m\sqrt{1+\ell}  A_{\rho}=0,\label{A2}\\ 
&&\coth\rho\left(\frac{\mathrm{d} A_{\rho}}{\mathrm{d} x^+}-\frac{\mathrm{d} A_{+}}{\mathrm{d} \rho}\right) +m\sqrt{1+\ell}  A_-=0. \label{A3}
\end{eqnarray}
By using Eq. (\ref{A2}), we get
\begin{equation}
	A_\rho=\frac{1}{m\sqrt{1+\ell} \cosh\rho\sinh\rho}\partial_{[+}A_{-]}.
\end{equation}
With the help of the transformation of the coordinate as before, Eqs. \eqref{A1} and \eqref{A3} can be recast as a set of coupled first order equations
\begin{eqnarray}\label{coupled}
&&4\,m\, z \frac{\mathrm{d} A_1(z)}{dz} - \sqrt{1 +\ell} \left[k^2 - \frac{m^2(z + 1)}{z - 1} - \omega^2 \right]A_1(z) - \sqrt{1 +\ell}\left[(k+\omega)^2 + m^2\right] A_2(z) = 0, \notag \\ 
&&4\,m\, z \frac{\mathrm{d} A_2(z)}{dz} + \sqrt{1 + \ell} \left[k^2 - \frac{m^2(z + 1)}{z - 1} - \omega^2 \right]A_2(z) + \sqrt{1 + \ell}\left[(k-\omega)^2 + m^2\right] A_1(z) = 0,
\end{eqnarray}
with $A_{1}=A_+ + A_-$ and $A_{2}=A_+ - A_-$. From Eq.~\eqref{coupled}, we can derive the corresponding second-order differential equations
\begin{eqnarray}\label{maxwelleq}
	z(1-z)\frac{\mathrm{d}^2A_{i}}{\mathrm{d} z^2}+(1-z)\frac{\mathrm{d}A_{i}}{\mathrm{d} z}+\left[\frac{k_+^2(1+\ell)}{4z}
	-\frac{k_-^2(1+\ell)}{4}-\frac{m^2(1+\ell)+2\varepsilon_i m\sqrt{1+\ell}}{4(1-z)}\right]A_{i}=0,
\end{eqnarray}
with $i = 1$ or $i = 2$. Interestingly, we can recover the scalar radial Eq.~\eqref{eq:radial}	by setting $R(z) \to A_{i}(z)$ and the scalar mass squared $m^2 \to m^2+2\varepsilon_i m/\sqrt{1+\ell}$ with $\varepsilon_1=-\varepsilon_2=1$. 

\begin{figure}[htbp]
	\subfigure{\includegraphics[scale=0.55]{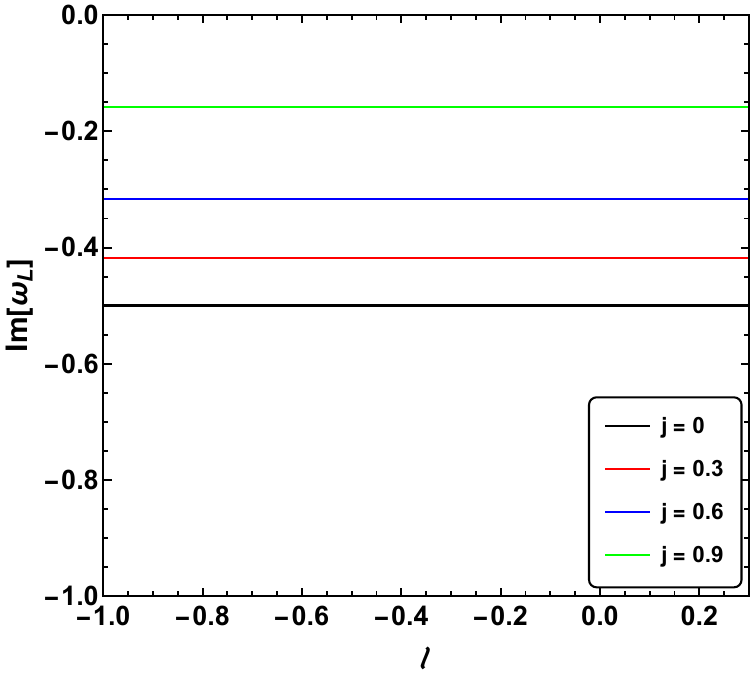}}\hspace{8ex}
	\subfigure{\includegraphics[scale=0.55]{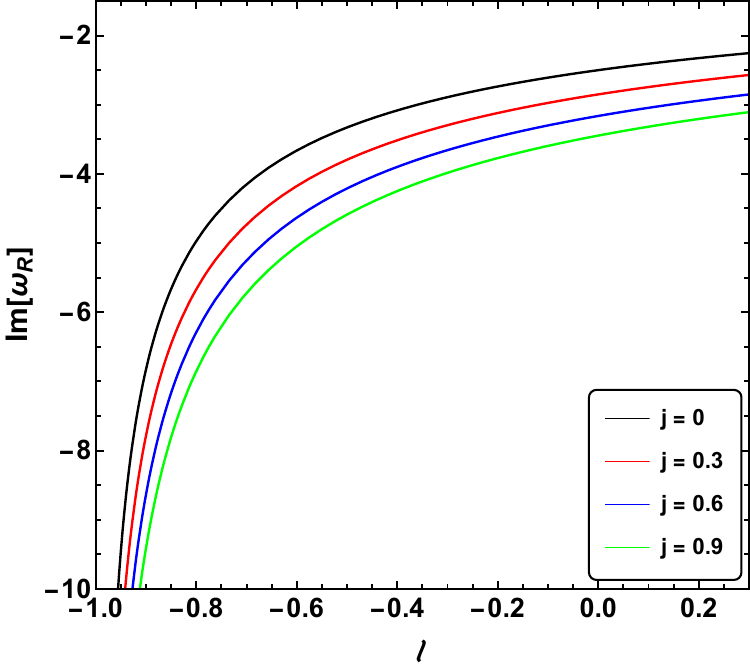}}\\ \vspace*{4ex}
	\subfigure{\includegraphics[scale=0.55]{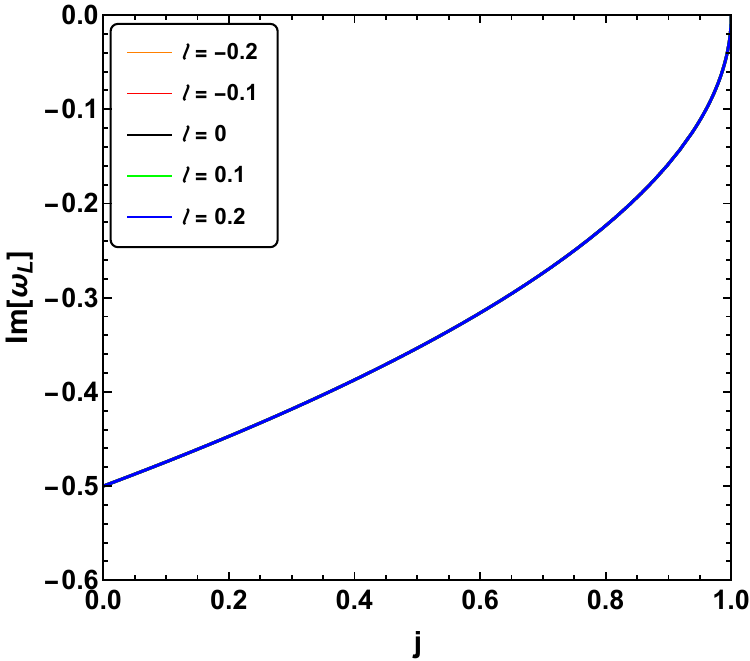}} \hspace{8ex}
	\subfigure{\includegraphics[scale=0.55]{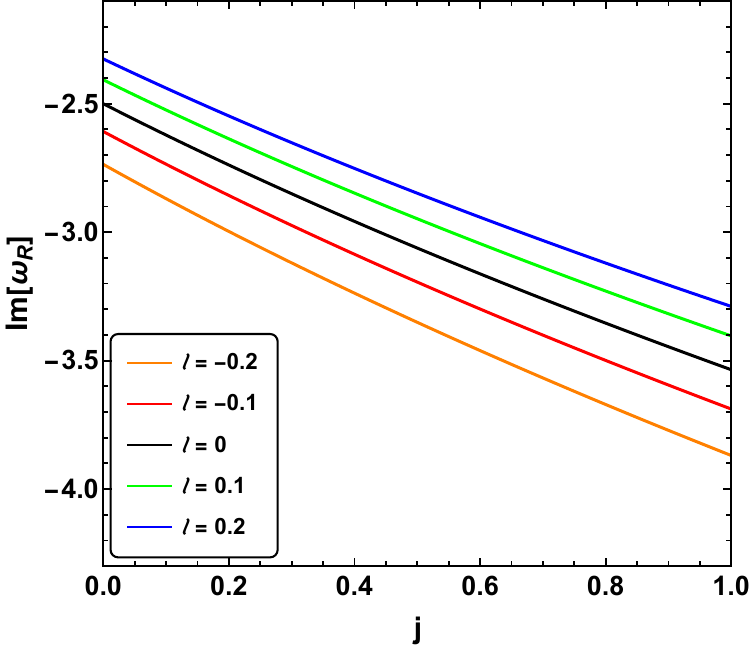}}
	\caption{ Variation of the imaginary parts of left- and right-moving quasinormal frequencies for the fundamental modes ($n=0$) under the vector perturbation with the LSB parameter $\ell$ and the rotation parameter $j$. Here we set $M=1$ and $m=0.5$.}\label{maxwellQNMs_s}
\end{figure}

As in the scalar case, at the horizon ($z = 0$), for the vector perturbation the solutions with the ingoing flux  are given by
\begin{eqnarray}\label{v5}
A_1&=&e_1\;z^{\alpha_v}(1-z)^{\beta_v+1}F(a_v+1,b_v+1;c_v;z), \notag\\
A_2&=&e_2\;z^{\alpha_v}(1-z)^{\beta_v}F(a_v,b_v;c_v;z),
\end{eqnarray}
with
\begin{eqnarray}
\alpha_v&=&-\frac{ik_+\sqrt{1+\ell}}{2},\qquad \beta_v=\frac{m\sqrt{1+\ell}}{2},\qquad c_v=1+2\alpha_v,\notag\\
a_v&=&\frac{(k_+-k_-)\sqrt{1+\ell}}{2i}+\beta_v,\qquad
b_v=\frac{(k_++k_-)\sqrt{1+\ell}}{2i}+\beta_v.
\end{eqnarray}
Eq.~\eqref{coupled} implies that the two coefficients $e_1$ and $e_2$ must satisfy
\begin{equation}\label{v7}
\frac{e_2}{e_1}=\frac{i(k_+-k_-)+m}{i(k_++k_-)-m}
= \frac{c_v-b_v-1}{b_v}.
\end{equation}
For the boundary condition at the infinity, it is reasonable to require that the energy flux vanishes, which is equivalent to
\begin{eqnarray}
\mathcal {F}_e\simeq (ik A_++i\omega A_-) A_+^*+c.c.\propto \left(A_1 A_2^*-A_1^*A_2 \right)=0,
\end{eqnarray}
where $c.c.$ stands for complex conjugate of the preceding terms. Expanding the energy flux near the asymptotic infinity $z \to 1$, we find that, for positive $m$, the leading-order term is proportional to
\begin{equation}
\left|\dfrac{\Gamma(c_v)\Gamma(a_v+b_v+2-c_v)}{\Gamma(a_v+1)\Gamma(b_v)}\right|^2(1-z)^{1-2\beta_v}.
\end{equation}
Thus, the flux boundary condition gives the following relation
\begin{equation}\label{vectorcase1}
a_v+1=-n ~~ \textrm{or} ~~ b_v=-n,
\end{equation}
from which we have
\begin{eqnarray}\label{maxwellqnms1}
\frac{i}{2}(k_+-k_-)\sqrt{1+\ell}=n+1+\frac{ m\sqrt{1+\ell}}{2},
\end{eqnarray}
or
\begin{eqnarray}\label{maxwellqnms2}
\frac{i}{2}(k_++k_-)\sqrt{1+\ell}=n+\frac{ m\sqrt{1+\ell}}{2}.
\end{eqnarray}
Then we can obtain the left- and right-moving quasinormal frequencies $\omega_L$ and $\omega_R$ for the vector perturbation
\begin{eqnarray}\label{MaxwellDLR}
\omega_L&=&k-\frac{2i(r_+-r_-)}{\sqrt{1+\ell}}\left(n+\frac{m\sqrt{1+\ell} }{2}\right),\\
\omega_R&=&-k-\frac{2i(r_++r_-)}{\sqrt{1+\ell}}\left(n+1+\frac{m\sqrt{1+\ell} }{2}\right).
\end{eqnarray}
In a similar manner to the first case, for negative $m$, we obtain a second set of relation $b_v+1=-n$ or $a_v=-n $, which leads to
\begin{eqnarray}
\omega_L&=&k-\frac{2i(r_+-r_-)}{\sqrt{1+\ell}}\left(n+1-\frac{m\sqrt{1+\ell} }{2}\right), \label{maxL}   \\
\omega_R&=&-k-\frac{2i(r_++r_-)}{\sqrt{1+\ell}}\left(n-\frac{m\sqrt{1+\ell} }{2}\right)\label{MaxwellNLR}.
\end{eqnarray}
Taking $\ell=0$, we can recover the quasinormal frequencies of the vector field in the BTZ black hole spacetime \cite{Birmingham:2001pj}. In analogy with the scalar and fermionic perturbations, the real parts of vector quasinormal frequencies are determined solely by the angular quantum number $k$. However, the imaginary parts of vector quasinormal frequencies depend on the LSB parameter $\ell$ and the rotation parameter $j$, except for $\omega_L$ with positive $m$ and $\omega_R$ with negative $m$ for the fundamental ($n=0$) modes where the imaginary parts are independent of the parameter $\ell$, just as shown in Fig. \ref{maxwellQNMs_s}. This is a key difference from the scalar and fermionic fields. For higher overtone modes, we observe that, regardless of the left branch or the right branch, $\mathrm{Im}[\omega_{L,R}]$ increases by increasing $\ell$, which is similar to the effect of $\ell$ on the scalar and fermionic QNMs.

Similar to the scalar and fermionic perturbations, according to the AdS/CFT correspondence, we obtain the conformal weights of the massive vector perturbation from Eq. (\ref{QNMsCFT}) and Eqs. (\ref{MaxwellDLR})---(\ref{MaxwellNLR}). So we get for positive $m$ 
\begin{eqnarray}
&&h_R^v = 1 + \frac{1}{2}m\sqrt{1+\ell},\qquad
h_L^v = \frac{1}{2}m\sqrt{1+\ell}, \notag\\
&&h_R^v + h_L^v = 1+m\sqrt{1+\ell}, \qquad
h_R^v - h_L^v = 1,
\end{eqnarray}
and for negative $m$
\begin{eqnarray}
&&h_R^v = -\frac{1}{2}m\sqrt{1+\ell},\qquad
h_L^v =1 -\frac{1}{2}m\sqrt{1+\ell}, \notag\\
&&h_R^v + h_L^v = 1-m\sqrt{1+\ell}, \qquad
h_R^v - h_L^v = -1,
\end{eqnarray}
which lead to the conformal dimension $\Delta=1\pm m\sqrt{1+\ell}$, implying that Eq. (\ref{HspinLR}) still remains valid for a vector perturbation around the rotating BTZ-like black hole in the Einstein-bumblebee gravity. Furthermore, we see that the characteristic relaxation time $\tau$ for a vector perturbation increases as the LSB parameter $\ell$ increases for $\omega_R$ with positive $m$ and $\omega_L$ with negative $m$ but is independent of $\ell$ for $\omega_L$ with positive $m$ and $\omega_R$ with negative $m$, which is different from the finding in the scalar and fermionic cases.

\section{Conclusions}

In this work we have obtained exact expressions for the QNMs of various perturbations, including the massive scalar, fermionic and vector perturbations, around a rotating BTZ-like black hole in the Einstein-bumblebee gravity. Imposing the vanishing energy flux boundary conditions at the AdS boundary, we found that, for various spins $s$, the LSB parameter $\ell$ leaves its imprint only on the imaginary parts of the quasinormal frequencies. We observed that the imaginary parts increase with the increase of $\ell$, which implies that the corresponding perturbation field decays more slowly for a larger $\ell$, except for the left-moving quasinormal frequencies $\omega_L$ with positive mass $m$ and the right-moving ones $\omega_R$ with negative mass $m$ for the fundamental ($n=0$) modes under the vector perturbation where the imaginary parts are independent of the parameter $\ell$. We also noted that, as the rotation parameter $j$ increases, the imaginary parts increase for $\omega_L$ but decrease for $\omega_R$. However, different from the imaginary parts, the real parts of quasinormal frequencies are independent of $\ell$ but depend on the angular quantum number $k$ regardless of the kind of the perturbations, which are the same as those in the standard BTZ black hole.

Furthermore, according to the AdS/CFT correspondence, we derived the left and right conformal weights ($h_L,h_R$) of the boundary operators dual to the scalar, fermionic and vector fields, and found that the expected universal relation $h_R + h_L = \Delta, ~h_R - h_L = \pm s$ still holds even for the BTZ-like black hole in the Einstein-bumblebee gravity, where the conformal dimension $\Delta$ is determined by the LSB parameter $\ell$ and the mass $m$ of various fields, i.e., $\Delta = 1 \pm \sqrt{1 + m^2(1+\ell)}$ for the scalar field ($s=0$), and $\Delta=1\pm m\sqrt{1+\ell}$ for the fermionic ($s=1/2$) and vector ($s=1$) fields. In addition, using the AdS/CFT dictionary, we observed that the characteristic relaxation time  $\tau$ increases as $\ell$ increases both for the left- and right-moving modes, which indicates that it will take more time for the perturbations to return to thermal equilibrium for a larger $\ell$, except for the cases of $\omega_L$ with positive $m$ and $\omega_R$ with negative $m$ for the vector perturbation. Our results give strong support to the AdS/CFT correspondence and could be helpful to understand the Einstein-bumblebee gravity with the Lorentz symmetry violation.

\begin{acknowledgments}		
	
This work was supported by the National Natural Science Foundation of China (Grant Nos. 12275079, 12547143 and 12035005), the National Key Research and Development Program of China (Grant No. 2020YFC2201400) and the innovative research group of Hunan Province (Grant No. 2024JJ1006).
	
\end{acknowledgments}

\bibliography{reference}

\end{document}